\documentclass[english,preprint,aps,pra,superscriptaddress]{revtex4-1}
\usepackage[T1]{fontenc}
\usepackage[latin9]{inputenc}
\usepackage{xcolor}
\usepackage{textcomp}
\usepackage{amsmath}
\usepackage{amssymb}
\usepackage{graphicx}
\usepackage{wasysym}
\PassOptionsToPackage{normalem}{ulem}
\usepackage{ulem}

\makeatletter

\providecommand{\tabularnewline}{\\}
\providecolor{lyxadded}{rgb}{0,0,1}
\providecolor{lyxdeleted}{rgb}{1,0,0}
\DeclareRobustCommand{\lyxsout}[1]{\ifx\\#1\else\sout{#1}\fi}

\makeatother

\usepackage{babel}
\begin{document}
\title{On the Proper Derivation of the Floquet-based Quantum Classical Liouville
Equation and Surface Hopping Describing a Molecule or Material Subject
to an External Field}
\author{Hsing-Ta Chen}
\email{hsingc@sas.upenn.edu}

\affiliation{Department of Chemistry, University of Pennsylvania, Philadelphia,
Pennsylvania 19104, U.S.A.}
\author{Zeyu Zhou}
\affiliation{Department of Chemistry, University of Pennsylvania, Philadelphia,
Pennsylvania 19104, U.S.A.}
\author{Joseph E. Subotnik}
\affiliation{Department of Chemistry, University of Pennsylvania, Philadelphia,
Pennsylvania 19104, U.S.A.}
\begin{abstract}
We investigate different approaches to derive the proper Floquet-based
quantum-classical Liouville equation (F-QCLE) for laser-driven electron-nuclear
dynamics. The first approach projects the operator form of the standard
QCLE onto the diabatic Floquet basis, then transforms to the adiabatic
representation. The second approach directly projects the QCLE onto
the Floquet adiabatic basis. Both approaches yield a form which is
similar to the usual QCLE with two modifications: 1. The electronic
degrees of freedom are expanded to infinite dimension. 2. The nuclear
motion follows Floquet quasi-energy surfaces. However, the second
approach includes an additional cross derivative force due to the
dual dependence on time and nuclear motion of the Floquet adiabatic
states. Our analysis and numerical tests indicate that this cross
derivative force is a fictitious artifact, suggesting that one cannot
safely exchange the order of Floquet state projection with adiabatic
transformation. Our results are in accord with similar findings by
Izmaylov \emph{et al}. (Ref.~41), who found that transforming to
the adiabatic representation must always be the last operation applied,
though now we have extended this result to a time-dependent Hamiltonian.
This paper and the proper derivation of the F-QCLE should lay the
basis for further improvements of Floquet surface hopping.
\end{abstract}
\maketitle

\section{Introduction\label{sec:Introduction}}

A computational understanding light-matter interactions for a molecular
system in a laser field is useful for interpreting spectroscopy and
photochemistry, where the dynamical interplay between electronic non-adiabatic
transitions and photon excitation plays an important role for many
exciting phenomena, such as molecular photodissociation\citep{regan_ultraviolet_1999,franks_orientation_1999,hilsabeck_photon_2019,hilsabeck_nonresonant_2019}
and coherent X-ray diffraction\citep{glownia_self-referenced_2016,glownia_glownia_2017,lemke_coherent_2017,fuller_drop--demand_2017}.
These phenomena usually involve dynamical processes in which electrons
in a molecular system can make a transition through either (a) non-adiabatic
coupling associated with the reorganization of nuclear configurations
or (b) radiative coupling in conjunction with absorption or emission
of photons. Thus, simulating these processes concurrently requires
accurate theoretical treatments of both non-adiabatic molecular dynamics
and light-matter interactions\citep{bajo_interplay_2014,hoffmann_light-matter_2018,abedi_exact_2010}.
Over the past decades, many successful simulation schemes have been
developed based on mixed quantum\textminus classical frameworks in
which the electronic wavefunction evolves according to quantum mechanics
while the nuclear degrees of freedom and the laser excitation are
treated as classical parameters in a time-dependent electronic Hamiltonian\citep{zhou_nonadiabatic_2020-1,richter_sharc:_2011,mitric_laser-field-induced_2009,mai_nonadiabatic_2018,thachuk_semiclassical_1996,bajo_mixed_2012,lisinetskaya_simulation_2011}.

Among the myriad of semiclassical dynamics, Floquet-based fewest
switch surface hopping (F-FSSH) has emerged as one of the most powerful
methods especially for simulating photodissociation and ionization
in a monochromatic laser field\citep{fiedlschuster_floquet_2016,fiedlschuster_surface_2017,horenko_theoretical_2001}.
In a nutshell, F-FSSH integrates Floquet theory with Tully's FSSH
algorithm\citep{tully_perspective:_2012}. The general idea is to
expand the electronic wavefunction in a Floquet state basis (with
the electronic states dressed by $e^{im\omega t}$ for an integer
$m$ and the laser frequency $\omega$), so that one can recast an
explicitly time-dependent Hamiltonian into a time-independent Floquet
Hamiltonian, albeit of infinite dimension. With the Floquet Hamiltonian,
one can simply employ Tully's FSSH method in the Floquet state representation
with a minimal modification\citep{fiedlschuster_floquet_2016,fiedlschuster_surface_2017}.
In addition to the standard advantages of the usual surface hopping
algorithm\textemdash stability, efficiency, and ease of incorporation
with electronic structure calculations\textemdash F-FSSH also yield
a better estimate for both electronic and nuclear observables than
other FSSH-based methods relying on instantaneous adiabatic surfaces\citep{zhou_nonadiabatic_2020-1}.
Furthermore, given the time-independent nature of the Floquet Hamiltonian,
many techniques designed to improve standard FSSH method, such as
velocity reversal and decoherence\citep{bittner_quantum_1995,jasper_electronic_2005,jasper_improved_2003,nelson_nonadiabatic_2013,subotnik_understanding_2016,jain_efficient_2016,fang_improvement_1999,bedard-hearn_mean-field_2005},
should be applicable within the Floquet formalism. That being said,
due to the fact that one cannot directly derive Tully's FSSH\textemdash but
rather only indirectly connect the equations of motion for FSSH dynamics
with the quantum-classical Liouville equation (QCLE) in the adiabatic
basis\citep{subotnik_can_2013,kapral_surface_2016}\textemdash a proper
understanding of the correct F-QCLE is crucial if we aim to make future
progress in semiclassical modeling of light-matter interactions. Moreover,
since it is well known that (1) FSSH performs best in an adiabatic
(rather than diabatic) basis\citep{tully_perspective:_2012,jasper_introductory_2004},
and (2) FSSH can be connected to the QCLE only in an adiabatic basis\citep{subotnik_can_2013,kapral_surface_2016},
we emphasize that, for the sake of future progress, we require a proper
understanding of the F-QCLE\emph{ in an adiabatic basis}.

Unfortunately, even without a light field, the proper derivation of
the correct QCLE in an adiabatic basis is tricky\textemdash one can
find two different versions of the QCLE if one invokes slightly different
formal derivations. Following Kapral's approach\citep{kapral_progress_2006,kapral_mixed_1999},
the proper derivation of the standard QCLE includes two operations:
(i) Wigner transformation and (ii) projection onto the adiabatic electronic
state basis. First, one performs a partial Wigner transformation with
respect to the nuclear degrees of freedom to obtain the operator form
of the QCLE. Wigner transformation provides an exact framework to
interpret the full quantum density matrix in terms of the joint electronic-nuclear
probability density in the phase space of the nuclear configuration
while retaining the quantum operator character of the electronic subsystem.
Second, one projects the operator form of the QCLE onto the adiabatic
electronic states basis obtained by diagonalizing the electronic Hamiltonian.
This adiabatic representation allows the connection to electronic
structure calculations in a mixed quantum-classical sense\citep{wong_solvent-induced_2002,wong_dissipative_2002,webster_nonadiabatic_1991,kelly_efficient_2013,kim_all-atom_2012}.
This approach is called the Wigner-then-Adiabatic (WA) approach. As
Izmaylov and co-workers have shown\citep{ryabinkin_analysis_2014},
however, exchanging these two operations (the Adiabatic-then-Wigner
(AW) approach) leads to a different QCLE that cannot capture geometric
phase effects arising from a conical intersection (even though short
time dynamcis can sometimes be accurate\citep{ando_non-adiabatic_2002,ando_mixed_2003}).

With this background in mind, the proper derivation of the F-QCLE
is now even more challenging. In addition to the two operations above,
there is a third step: one needs to dress the electronic states and
expand the density matrix in the Floquet state basis. In the literature
to date, the F-QCLE has been derived via the AW approach (projecting
in the Floquet adiabatic representation, and then performing partial
Wigner transformation)\citep{horenko_theoretical_2001}. Nevertheless,
as shown by Izmaylov and co-workers\citep{ryabinkin_analysis_2014,kapral_mixed_1999},
even in the limit of a time-independent Hamiltonian, such an (incorrect)
AW approach will lead to a QCLE that neglects geometric phase related
features in the nuclear dynamics or introduces artificial nuclear
effects. Despite recent progress, the proper derivation of the F-QCLE
is still an open question.

In this paper, our goal is to explore different approaches to derive
the F-QCLE as we shuffle the three key operations and quantify their
differences in the context of driven electron-nuclear dynamics. By
isolating the correct F-QCLE, our work will not only validate F-FSSH
methods, it should also provide means to improve surface hopping methods.
This paper is organized as follows. In Sec.~\ref{sec:Three-operations},
we formulate the three operations that are required to derive the
F-QCLE. In Sec.~\ref{sec:Floquet-QCLE}, we derive F-QCLEs via approaches
with different ordering of operations. In Sec.~\ref{sec:Results:},
we implement F-FSSH calculations corresponding to these F-QCLEs and
analyze their results for a modified avoided crossing model. We conclude
with an outlook for the future in Sec.~\ref{sec:Conclusion}.

For notation, we denote a quantum operator by $\hat{H}$ and use bold
font for matrix $\mathbf{H}$. We use $\widetilde{\boldsymbol{H}}$
to denote the corresponding matrix in the expanded Floquet basis (infinite
dimensional). The nuclear position and momentum are $\vec{R}=\{R^{\alpha}\}$,
$\vec{P}=\{P^{\alpha}\}$ where $\alpha$ is the nuclear coordinate
index. We use a shorthand notation for dot product: $X^{\alpha}\cdot Y^{\alpha}=\sum_{\alpha}X^{\alpha}Y^{\alpha}$
.

\section{Three Operations\label{sec:Three-operations}}

In the context of driven electron-nuclear dynamics, let us formulate
the three necessary operations for deriving F-QCLE. For a coupled
electron-nuclei system driven by an external field of frequency $\omega$,
the total Hamiltonian takes the form of $\hat{H}=\hat{T}(\hat{P}^{\alpha})+\hat{V}(\hat{R}^{\alpha},t)$
where $\hat{T}(\hat{P}^{\alpha})=\sum_{\alpha}\frac{(\hat{P}^{\alpha})^{2}}{2M^{\alpha}}$
is the nuclear kinetic energy and $\hat{V}(\hat{R}^{\alpha},t)$ is
the electronic Hamiltonian with explicit time periodicity $\hat{V}(t)=\hat{V}(t+\tau)$
with $\tau=2\pi/\omega$. Formally, the dynamics of the total system
can be described by the time-dependent Schrödinger equation (TDSE)
$i\hbar\frac{\partial}{\partial t}|\Psi\rangle=\hat{H}|\Psi\rangle$
of the total wavefunction $|\Psi\rangle$ or the quantum Liouville
equation (QLE) $\frac{\partial}{\partial t}\hat{\rho}=-\frac{i}{\hbar}[\hat{H},\hat{\rho}]$
of the total density matrix $\hat{\rho}=|\Psi\rangle\langle\Psi|$.
To derive F-QCLE, we need to apply the following three operations
to the QLE.

\subsection{Partial Wigner transformation of the nuclear degrees of freedom}

To describe the dynamics in a mixed quantum-classical sense, we will
follow Kapral\textquoteright s approach and perform a partial Wigner
transformation with respect to the nuclear degrees of freedom
\begin{equation}
\hat{\rho}_{W}\left(\vec{R},\vec{P},t\right)=\frac{1}{(2\pi\hbar)^{N}}\int d\vec{S}\left\langle \vec{R}+\frac{\vec{S}}{2}\right|\hat{\rho}(t)\left|\vec{R}-\frac{\vec{S}}{2}\right\rangle e^{-i\vec{P}\cdot\vec{S}/\hbar}\label{eq:partial_Wigner_distribution}
\end{equation}
where $N$ is the dimension of the nuclear coordinate. A nuclear position
eigenstate is defined as $\hat{R}^{\alpha}|R^{\alpha}\rangle=R^{\alpha}|R^{\alpha}\rangle$.
In Eq.~\eqref{eq:partial_Wigner_distribution}, the density matrix
operator has been transformed into a Wigner wavepacket in phase space
with coordinates $(\vec{R},\vec{P})$. In what follows, we will denote
the partial Wigner transformed operator by the subscript $W$ {[}for
example $\hat{V}(\hat{R}^{\alpha},t)\rightarrow\hat{V}_{W}(R^{\alpha},t)${]}.
Note that, after the partial Wigner transformation, $\hat{\rho}_{W}$
and $\hat{V}_{W}$ remain electronic operators while $R^{\alpha}$
and $P^{\alpha}$ are parameters.

The equation of motion of the Wigner wavepacket can be obtained by
transforming the QLE by $\frac{\partial}{\partial t}\hat{\rho}_{W}=-\frac{i}{\hbar}[(\hat{H}\hat{\rho})_{W}-(\hat{\rho}\hat{H})_{W}]$.
The Wigner transform of operator products can be expanded further
by the Wigner\textendash Moyal operator $(\hat{H}\hat{\rho})_{W}=\hat{H}_{W}e^{-i\hbar\overleftrightarrow{\Lambda}/2}\hat{\rho}_{W}$
with $\overleftrightarrow{\Lambda}=\sum_{\alpha}\overleftarrow{\frac{\partial}{\partial P^{\alpha}}}\overrightarrow{\frac{\partial}{\partial R^{\alpha}}}-\overleftarrow{\frac{\partial}{\partial R^{\alpha}}}\overrightarrow{\frac{\partial}{\partial P^{\alpha}}}$\citep{kapral_mixed_1999}.
Then, if we expand the the Wigner\textendash Moyal operator in Taylor
series and truncate to the first order of $\hbar$, we obtain the
operator form of the QCLE
\begin{equation}
\frac{\partial}{\partial t}\hat{\rho}_{W}=-\frac{i}{\hbar}\left[\hat{V}_{W},\hat{\rho}_{W}\right]-\frac{P^{\alpha}}{M^{\alpha}}\frac{\partial\hat{\rho}_{W}}{\partial R^{\alpha}}+\frac{1}{2}\left\{ \frac{\partial\hat{V}_{W}}{\partial R^{\alpha}},\frac{\partial\hat{\rho}_{W}}{\partial P^{\alpha}}\right\} .\label{eq:QCLE-operator}
\end{equation}
Here, the commutator is $\left[\hat{A},\hat{B}\right]=\hat{A}\hat{B}-\hat{B}\hat{A}$
and the anti-commutator is $\left\{ \hat{A},\hat{B}\right\} =\hat{A}\hat{B}+\hat{B}\hat{A}$.
Note that Eq.~\eqref{eq:QCLE-operator} is exact if the partial Wigner
transformed Hamiltonian is quadratic in $R^{\alpha}$, for example
harmonic oscillators.

To propagate the Wigner wavepacket in Eq.~\eqref{eq:QCLE-operator},
one must project the operator form of the QCLE in an electronic basis.
One straightforward choice is to use a complete set of diabatic states
for the electronic subsystem $\{|\mu\rangle\}$; such a set does not
depend on any nuclear configuration. With this electronic diabatic
basis, one can derive equations of motion for the density matrix ($A_{\nu\mu}^{\text{dia}}=\langle\nu|\hat{\rho}_{W}|\mu\rangle$)
using matrix elements of the electronic Hamiltonian, $V_{\nu\mu}(\vec{R},t)=\langle\nu|\hat{V}_{W}(\vec{R},t)|\mu\rangle$.
However, for many realistic electron-nuclei systems (and certainly
any ab initio calculations), this diabatic QCLE cannot be solved since
finding a complete set of exactly diabatic electronic states over
a large set of nuclear geometries is rigorously impossible and quite
demanding in practice even for approximate diabats. 

\subsection{Dress the electronic basis in the Floquet formalism}

Let us now focus on the Floquet formalism, according to which one
solves the TDSE by transforming the time-dependent Hamiltonian into
a time-independent Floquet Hamiltonian in an extended Hilbert space
of infinite dimension. For the moment let us ignore all nuclear motion
and focus on the electronic subsystem exclusively. According to Floquet
theory, we utilize the time periodicity of the electronic Hamiltonian
and dress the electronic diabatic states $\{|\mu\rangle\}$ by a time-periodic
function $e^{im\omega t}$ where $m$ is an integer formally from
$-\infty$ to $\infty$. We denote the dressed state as the the \emph{Floquet
diabatic state} $|\mu m\rangle\equiv e^{im\omega t}|\mu\rangle$.
In terms of the Floquet diabatic basis, a time periodic electronic
wavefunction can be expressed as $|\Psi\rangle=\sum_{\mu m}\tilde{c}_{\mu m}|\mu m\rangle$
where $\tilde{c}_{\mu m}$ is an infinite dimensional state vector.
The electronic wavefunction coefficient must satisfy the electronic
TDSE
\begin{equation}
i\hbar\sum_{\mu m}\frac{\partial\tilde{c}_{\mu m}}{\partial t}|\mu m\rangle=\sum_{\mu m}\hat{V}^{\text{F}}(t)|\mu m\rangle\tilde{c}_{\mu m}\label{eq:TDSE-Floquet-operator}
\end{equation}
where the Floquet Hamiltonian operator is defined as
\begin{equation}
\hat{V}^{\text{F}}(t)\equiv\hat{V}(t)-i\hbar\frac{\partial}{\partial t}.\label{eq:define-Floquet_Hamiltonian}
\end{equation}
Next, we close Eq.~\ref{eq:TDSE-Floquet-operator} by multiplying
both sides by $\langle\nu|$ and write $\langle\nu|\hat{V}^{\text{F}}(t)|\mu m\rangle=\sum_{n}\widetilde{V}_{\nu n,\mu m}^{\text{F}}e^{in\omega t}$
as a Fourier series:
\begin{equation}
i\hbar\sum_{m}\frac{\partial\tilde{c}_{\nu m}}{\partial t}e^{im\omega t}=\sum_{\mu m}\sum_{n}\widetilde{V}_{\nu n,\mu m}^{\text{F}}e^{in\omega t}\tilde{c}_{\mu m}.
\end{equation}
Thus, the TDSE in Eq.~\eqref{eq:TDSE-Floquet-operator} can be solved
by grouping together all terms with the same time dependence, leading
to the equation of motion for $\tilde{c}$
\begin{equation}
i\hbar\frac{\partial}{\partial t}\tilde{c}_{\nu n}=\sum_{\mu m}\widetilde{V}_{\nu n,\mu m}^{\text{F}}\tilde{c}_{\mu m}.\label{eq:TDSE-Floquet-matrix}
\end{equation}
The matrix elements of the Floquet Hamiltonian can be obtained by
performing a Fourier transformation on the matrix elements 
\begin{equation}
\widetilde{V}_{\nu n,\mu m}^{\text{F}}=\langle\langle\nu n|\hat{V}^{\text{F}}|\mu m\rangle\rangle=\frac{1}{\tau}\int_{0}^{\tau}dt\left\langle \nu\right|\hat{V}^{\text{F}}(t)\left|\mu\right\rangle e^{i(m-n)\omega t}.\label{eq:double-bracket}
\end{equation}
Here, we define the double bracket projection of an electronic operator
by $\langle\langle\nu n|\cdots|\mu m\rangle\rangle=\frac{1}{\tau}\int_{0}^{\tau}dt\left\langle \nu\right|\cdots\left|\mu\right\rangle e^{i(m-n)\omega t}$.
\emph{Given that the electronic Hamiltonian operator $\hat{V}(t)$
is periodic in time, the double-bracket projection eliminates all
time dependence} and the Floquet Hamiltonian matrix reads
\begin{equation}
\widetilde{V}_{\nu n,\mu m}^{\text{F}}=\langle\langle\nu n|\hat{V}|\mu m\rangle\rangle+\delta_{\mu\nu}\delta_{mn}m\hbar\omega.\label{eq:Floquet_Hamiltonian_projection_notation}
\end{equation}
In the end, with this time-independent Hamiltonian, Eq.~\eqref{eq:TDSE-Floquet-matrix}
can be formally solved by the exponential operator $\exp(-i\widetilde{\mathbf{V}}^{\text{F}}t/\hbar)$
with an arbitrary initial state.

At this point, we will allow nuclei to move and turn out attention
to the equation of motion for the density matrix $\widehat{\rho}_{W}(\vec{R},\vec{P},t)$
within the Floquet diabatic basis. The Wigner-transformed density
matrix in the Floquet diabatic representation is
\begin{equation}
\widetilde{A}_{\nu n,\mu m}^{\text{dia}}(\vec{R},\vec{P},t)=\langle\nu n|\widehat{\rho}_{W}(\vec{R},\vec{P},t)|\mu m\rangle.
\end{equation}
For a proper F-QCLE, we will need to calculate the time derivative
of $\widetilde{\mathbf{A}}^{\text{dia}}$ 
\begin{equation}
\frac{\partial}{\partial t}\widetilde{A}_{\nu n,\mu m}^{\text{dia}}=\left\langle \nu n\right|\frac{\partial\widehat{\rho}_{W}}{\partial t}\left|\mu m\right\rangle -i\left(n-m\right)\hbar\omega\widetilde{A}_{\nu n,\mu m}^{\text{dia}}\label{eq:dAdt-diabatic}
\end{equation}
where the Floquet diabatic states depends on time explicitly. We begin
by using Eq.~\eqref{eq:QCLE-operator} to project $\frac{\partial}{\partial t}\hat{\rho}_{W}$
into a Floquet diabatic basis. For the commutator term in Eq.~\eqref{eq:QCLE-operator},
we can divide the operator product into matrix multiplication: 
\begin{eqnarray}
\langle\nu n|\left[\hat{V}_{W},\hat{\rho}_{W}\right]|\mu m\rangle & = & \sum_{\lambda l}\langle\langle\nu n|\hat{V}_{W}|\lambda l\rangle\rangle\widetilde{A}_{\lambda l,\mu m}^{\text{dia}}-\widetilde{A}_{\nu n,\lambda l}^{\text{dia}}\langle\langle\lambda l|\hat{V}_{W}|\mu m\rangle\rangle\nonumber \\
 & = & [\widetilde{\mathbf{V}}_{W},\widetilde{\mathbf{A}}^{\text{dia}}]_{\nu n,\mu m}.\label{eq:commutator-diabatic}
\end{eqnarray}
Here, we have inserted the identity of the diabatic electronic basis
($\hat{1}=\sum_{\lambda}|\lambda\rangle\langle\lambda|$) and expanded
the time-dependent coefficients in terms of a Fourier series; see
Appendix~\ref{sec:The-trick} for more details. Furthermore, if we
combine Eq.~\eqref{eq:commutator-diabatic} with the second term
on the RHS of Eq.~\eqref{eq:dAdt-diabatic}, we can write the sum
of both terms as $[\widetilde{\mathbf{V}}^{\text{F}},\widetilde{\mathbf{A}}^{\text{dia}}]$,
i.e. we can replace $\widetilde{\mathbf{V}}_{W}$ with $\widetilde{\mathbf{V}}^{\text{F}}$.

For the anti-commutator term, we can use the same procedure to divide
the operator product 
\begin{eqnarray}
\langle\nu n|\left\{ \frac{\partial\hat{V}_{W}}{\partial R^{\alpha}},\frac{\partial\hat{\rho}_{W}}{\partial P^{\alpha}}\right\} |\mu m\rangle & = & \sum_{\lambda l}\langle\langle\nu n|\frac{\partial\hat{V}_{W}}{\partial R^{\alpha}}|\lambda l\rangle\rangle\frac{\partial\widetilde{A}_{\lambda l,\mu m}^{\text{dia}}}{\partial P^{\alpha}}\nonumber \\
 &  & +\frac{\partial\widetilde{A}_{\nu n,\lambda l}^{\text{dia}}}{\partial P^{\alpha}}\langle\langle\lambda l|\frac{\partial\hat{V}_{W}}{\partial R^{\alpha}}|\mu m\rangle\rangle
\end{eqnarray}
where $\langle\nu n|\frac{\partial\widehat{\rho}_{W}}{\partial P^{\alpha}}|\mu m\rangle=\frac{\partial}{\partial P^{\alpha}}\widetilde{A}_{\nu n,\mu m}^{\text{dia}}$.
Note that, since the Floquet diabatic states do not depend on the
nuclear coordinate, we can rewrite the derivative of the electronic
Hamiltonian in terms of the Floquet Hamiltonian
\begin{equation}
\langle\langle\nu n|\frac{\partial\hat{V}_{W}}{\partial R^{\alpha}}|\mu m\rangle\rangle=\frac{\partial}{\partial R^{\alpha}}\widetilde{V}_{\nu n,\mu m}^{\text{F}}.\label{eq:dVdR-diabatic}
\end{equation}
 In the end, we may combine the above expressions to write down a
complete \emph{diabatic} F-QCLE
\begin{eqnarray}
\frac{\partial}{\partial t}\widetilde{\mathbf{A}}^{\text{dia}} & = & -\frac{i}{\hbar}\left[\widetilde{\mathbf{V}}^{\text{F}},\widetilde{\mathbf{A}}^{\text{dia}}\right]-\frac{P^{\alpha}}{M}\frac{\partial\widetilde{\mathbf{A}}^{\text{dia}}}{\partial R^{\alpha}}\nonumber \\
 &  & +\frac{1}{2}\left\{ \frac{\partial\widetilde{\mathbf{V}}^{\text{F}}}{\partial R^{\alpha}},\frac{\partial\widetilde{\mathbf{A}}^{\text{dia}}}{\partial P^{\alpha}}\right\} \label{eq:F-QCLE-diabatic}
\end{eqnarray}
As a final remark, we emphasize that the Floquet Hamiltonian $\widetilde{\mathbf{V}}^{\text{F}}=\widetilde{\mathbf{V}}^{\text{F}}(\vec{R})$
is a time-independent matrix, so Eq.~\eqref{eq:F-QCLE-diabatic}
is simply the diabatic QCLE corresponding to an infinitely large electronic
Hamiltonian $\widetilde{\mathbf{V}}^{\text{F}}$.

\subsection{Transformation to the adiabatic representation}

To recast the diabatic F-QCLE in an adiabatic representation, we
diagonalize the Floquet Hamiltonian matrix by solving the eigenvalue
problem: 
\begin{equation}
\sum_{\nu n}\widetilde{V}_{\mu m,\nu n}^{\text{F}}(\vec{R})G_{\nu n}^{J}(\vec{R})={\cal E}_{J}^{\text{F}}(\vec{R})G_{\mu m}^{J}(\vec{R}).\label{eq:Floquet_adiabatic_eigenvalue}
\end{equation}
The eigenvalues ${\cal E}_{J}^{\text{F}}={\cal E}_{J}^{\text{F}}(\vec{R})$
are the so-called Floquet quasi-energies with corresponding eigenvectors
$G_{\mu m}^{J}(\vec{R}).$ Since $\widetilde{\mathbf{V}}^{\text{F}}$
is Hermitian, we can choose the eigenvectors $G_{\mu m}^{J}$ to be
othornormal so that we have the identities $\mathbf{G}^{\dagger}\mathbf{G}=\mathbf{G}\mathbf{G}^{\dagger}=\mathbf{I}$,
i.e. $\sum_{\lambda\ell}G_{\lambda\ell}^{J*}G_{\lambda\ell}^{K}=\delta_{JK}$
and $\sum_{L}G_{\mu m}^{L}G_{\nu n}^{L*}=\delta_{\mu m,\nu n}$. The
Floquet adiabatic state corresponding to the quasi-energy ${\cal E}_{J}^{\text{F}}$
are 
\begin{equation}
|\Phi^{J}(\vec{R},t)\rangle=\sum_{\mu m}G_{\mu m}^{J}(\vec{R})\left|\mu m\right\rangle .\label{eq:Floquet_adiabatic_eigenstate}
\end{equation}
As a practical matter, although $\widetilde{\mathbf{V}}^{\text{F}}$
is infinite dimensional, we can truncate highly oscillating Floquet
states by replacing $\sum_{m=-\infty}^{\infty}$ with $\sum_{m=-m_{\text{max}}}^{m_{\text{max}}}$.

With this Floquet adiabatic state basis, the probability density can
be obtained by a diabatic-to-adiabatic transformation 
\begin{equation}
\widetilde{A}_{JK}^{\text{adi}}(\vec{R},\vec{P},t)=\langle\Phi^{J}|\widehat{\rho}_{W}\left(\vec{R},\vec{P},t\right)|\Phi^{K}\rangle=\left(\mathbf{G}^{\dagger}\widetilde{\mathbf{A}}^{\text{dia}}\mathbf{G}\right)_{JK}
\end{equation}
in the Floquet adiabatic representation. Note that, since the eigenvectors
$G_{\mu m}^{J}(\vec{R})$ do not depend on time explicitly, the time-derivative
of the adiabatic probability density can be calculated simply to be:
\begin{equation}
\frac{\partial\widetilde{\mathbf{A}}^{\text{adi}}}{\partial t}=\mathbf{G}^{\dagger}\frac{\partial\widetilde{\mathbf{A}}^{\text{dia}}}{\partial t}\mathbf{G}.\label{eq:dAdt-adiabatic}
\end{equation}

We are now ready to derive the adiabatic F-QCLE in the following section.

\section{Different Approaches to derive F-QCLE\label{sec:Floquet-QCLE}}

In this section, we present several approaches with different orders
for the three operations above; as will be shown, different orders
will result in different adiabatic F-QCLEs. We summarize these approaches
and the corresponding F-QCLEs in Table~\ref{tab:F-QCLEs-via-different}.

\subsection{Wigner-Floquet-Adiabatic (WFA) approach}

Our first approach follows the order presented above: we first perform
partial Wigner transformation, we second project to a Floquet diabatic
basis, we third transform to an adiabatic representation. For the
last step, following Eq.~\eqref{eq:dAdt-adiabatic}, we transform
the diabatic F-QCLE by sandwiching the diabatic F-QCLE (Eq.~\eqref{eq:F-QCLE-diabatic})
with $\mathbf{G}^{\dagger}$ and $\mathbf{G}$. The first term on
the right hand side of Eq.~\eqref{eq:F-QCLE-diabatic} (the commutator
term) becomes
\begin{eqnarray}
\sum_{\nu n}\sum_{\mu m}G_{\nu n}^{J*}\left[\widetilde{V}^{\text{F}},\widetilde{A}^{\text{dia}}\right]_{\nu n,\mu m}G_{\mu m}^{K} & = & \left({\cal E}_{J}^{\text{F}}-{\cal E}_{K}^{\text{F}}\right)\widetilde{A}_{JK}^{\text{adi}}
\end{eqnarray}
For the second term on the right hand side of Eq.~\eqref{eq:F-QCLE-diabatic},
since the Floquet adiabatic states depend on the nuclear coordinate
$\vec{R}$, the $R^{\alpha}$ derivative of the density must yield
\begin{equation}
\sum_{\nu n}\sum_{\mu m}G_{\nu n}^{J*}\frac{\partial\widetilde{A}_{\nu n,\mu m}^{\text{dia}}}{\partial R^{\alpha}}G_{\mu m}^{K}=\frac{\partial\widetilde{A}_{JK}^{\text{adi}}}{\partial R^{\alpha}}+\sum_{L}\left(D_{JL}^{\alpha}\widetilde{A}_{LK}^{\text{adi}}-\widetilde{A}_{JL}^{\text{adi}}D_{LK}^{\alpha}\right)
\end{equation}
where the derivative coupling is $D_{JK}^{\alpha}=\langle\langle\Phi^{J}|\frac{\partial}{\partial R^{\alpha}}|\Phi^{K}\rangle\rangle=\sum_{\mu m}G_{\mu m}^{J*}\frac{\partial G_{\mu m}^{K}}{\partial R^{\alpha}}$
corresponding to the change of the Floquet adiabatic states with respect
to the nuclear coordinate $R^{\alpha}$. Note that, if the Floquet
Hamiltonian is real-valued, the diagonal element of the derivative
coupling is zero ($D_{JJ}^{\alpha}=0$). For the third term on the
right hand side of Eq.~\eqref{eq:F-QCLE-diabatic} (the anti-commutator
term), the $R^{\alpha}$ derivative of the Floquet Hamiltonian can
be written in terms of the force matrix 
\begin{equation}
\sum_{\nu n}\sum_{\mu m}G_{\nu n}^{J*}\frac{\partial\widetilde{V}_{\nu n,\mu m}^{\text{F}}}{\partial R^{\alpha}}G_{\mu m}^{K}=-F_{JK}^{\alpha}\label{eq:Force-matrix}
\end{equation}
explicitly,
\[
F_{JK}^{\alpha}=-\frac{\partial{\cal E}_{J}^{\text{F}}}{\partial R^{\alpha}}\delta_{JK}+({\cal E}_{J}^{\text{F}}-{\cal E}_{K}^{\text{F}})D_{JK}^{\alpha}.
\]
The force matrix accounts for direct changes in the nuclear momentum
associated with the electronic coupling. One can understand the diagonal
element $F_{JJ}^{\alpha}=-\frac{\partial{\cal E}_{J}^{\text{F}}}{\partial R^{\alpha}}$
as the classical force for nuclear dynamics moving along the $J$-th
Floquet quasi-energy surface in the phase space. The off-diagonal
force term $F_{JK}^{\alpha}=({\cal E}_{J}^{\text{F}}-{\cal E}_{K}^{\text{F}})D_{JK}^{\alpha}$
for $J\neq K$ corresponds to nuclear velocity rescaling associated
with non-adiabatic transitions between Floquet quasi-energy surfaces\citep{kapral_mixed_1999,kapral_progress_2006}.

Finally, the F-QCLE via the WFA approach reads 
\begin{eqnarray}
\frac{\partial}{\partial t}\widetilde{A}_{JK}^{\text{adi}} & = & -\frac{i}{\hbar}\left({\cal E}_{J}^{\text{F}}-{\cal E}_{K}^{\text{F}}\right)\widetilde{A}_{JK}^{\text{adi}}\nonumber \\
 &  & -\frac{P^{\alpha}}{M^{\alpha}}\frac{\partial\widetilde{A}_{JK}^{\text{adi}}}{\partial R^{\alpha}}-\frac{P^{\alpha}}{M^{\alpha}}\sum_{L}\left(D_{JL}^{\alpha}\widetilde{A}_{LK}^{\text{adi}}-\widetilde{A}_{JL}^{\text{adi}}D_{LK}^{\alpha}\right)\nonumber \\
 &  & -\frac{1}{2}\sum_{L}\left(F_{JL}^{\alpha}\frac{\partial\widetilde{A}_{LK}^{\text{adi}}}{\partial P^{\alpha}}+\frac{\partial\widetilde{A}_{JL}^{\text{adi}}}{\partial P^{\alpha}}F_{LK}^{\alpha}\right)\label{eq:F-QCLE-WDA}
\end{eqnarray}
We find that Eq.~\ref{eq:F-QCLE-WDA} takes exactly the same form
as the standard QCLE in the adiabatic representation (for electron-nuclear
dynamics without a driving laser).

\subsection{Wigner-Adiabatic-Floquet (WAF) approach}

For the second approach, we exchange the ``adiabatic'' and ``Floquet''
operations after the partial Wigner transformation. In this case,
we directly project Eq.~\eqref{eq:QCLE-operator} onto the Floquet
adiabatic basis $|\phi^{J}(\vec{R},t)\rangle$. Namely, we make the
diabatic-to-adiabatic transform of the Floquet electronic basis prior
to the projection onto the dressed electronic states. Thus, we consider
this path as the \emph{Wigner-Adiabatic-Floquet} (WAF) approach.

Overall, we apply a procedure similar to what was used in Eq.~\ref{eq:F-QCLE-WDA}.
For the commutator term, we divide operator products using the same
technique as in Appendix~\ref{sec:The-trick}. The $R^{\alpha}$
derivative term yields a derivative coupling term as the Floquet adiabatic
basis depends on $R^{\alpha}$ explicitly. In the end, the WAF approach
includes the first three terms exactly as Eq.~\eqref{eq:F-QCLE-WDA}.
However, from the anti-commutator term of Eq.~\eqref{eq:QCLE-operator},
the WAF approach leads to an additional cross derivative force. To
see this, we focus on the derivative of the electronic Hamiltonian
in the adiabatic representation 
\begin{equation}
\langle\langle\Phi^{J}|\frac{\partial\hat{V}}{\partial R^{\alpha}}|\Phi^{K}\rangle\rangle=\langle\langle\Phi^{J}|\frac{\partial\hat{V}^{\text{F}}}{\partial R^{\alpha}}|\Phi^{K}\rangle\rangle+i\hbar\langle\langle\Phi^{J}|\frac{\partial}{\partial R^{\alpha}}\frac{\partial}{\partial t}|\Phi^{K}\rangle\rangle
\end{equation}
where we have used the definition of the Floquet Hamiltonian $\frac{\partial\hat{V}^{\text{F}}}{\partial R^{\alpha}}=\frac{\partial\hat{V}}{\partial R^{\alpha}}-i\hbar\frac{\partial}{\partial R^{\alpha}}\frac{\partial}{\partial t}$.
The derivative of the electronic Hamiltonian yields two terms: first,
the same force matrix we obtained in Eq.~\eqref{eq:Force-matrix}:
\begin{equation}
\langle\langle\Phi^{J}|\frac{\partial\hat{V}^{\text{F}}}{\partial R^{\alpha}}|\Phi^{K}\rangle\rangle=-F_{JK}^{\alpha}
\end{equation}
second, a cross derivative force comes from the explicit dependence
of the Floquet adiabatic states on both the nuclear coordinates and
time
\[
i\hbar\langle\langle\Phi^{J}|\frac{\partial}{\partial R^{\alpha}}\frac{\partial}{\partial t}|\Phi^{K}\rangle\rangle=-\chi_{JK}^{\alpha},
\]
and explicitly, 
\begin{equation}
\chi_{JK}^{\alpha}=\sum_{\mu m}m\hbar\omega G_{\mu m}^{J*}\frac{\partial G_{\mu m}^{K}}{\partial R^{\alpha}}.\label{eq:cross-derivative-force}
\end{equation}
Note that, unlike the derivative coupling $D_{JK}^{\alpha}$, the
diagonal element of $\chi_{JK}^{\alpha}$ is non-zero and real-valued.

Finally, in the same Floquet adiabatic basis as given by Eq.~\eqref{eq:Floquet_adiabatic_eigenstate},
the F-QCLE reads:
\begin{eqnarray}
\frac{\partial}{\partial t}\widetilde{A}_{JK}^{\text{adi}} & = & -\frac{i}{\hbar}\left({\cal E}_{J}^{\text{F}}-{\cal E}_{K}^{\text{F}}\right)\widetilde{A}_{JK}^{\text{adi}}\nonumber \\
 &  & -\frac{P^{\alpha}}{M^{\alpha}}\frac{\partial\widetilde{A}_{JK}^{\text{adi}}}{\partial R^{\alpha}}-\frac{P^{\alpha}}{M^{\alpha}}\sum_{L}\left(D_{JL}^{\alpha}\widetilde{A}_{LK}^{\text{adi}}-\widetilde{A}_{JL}^{\text{adi}}D_{LK}^{\alpha}\right)\nonumber \\
 &  & -\frac{1}{2}\sum_{L}\left(F_{JL}^{\alpha}+\chi_{LJ}^{\alpha*}\right)\frac{\partial\widetilde{A}_{LK}^{\text{adi}}}{\partial P^{\alpha}}+\frac{\partial\widetilde{A}_{JL}^{\text{adi}}}{\partial P^{\alpha}}\left(F_{LK}^{\alpha}+\chi_{LK}^{\alpha}\right)\label{eq:F-QCLE-WAD}
\end{eqnarray}
We observe that, while Eq.~\eqref{eq:F-QCLE-WDA} and Eq.~\eqref{eq:F-QCLE-WAD}
take the same form, the ``effective'' force matrix (defined as the
coefficients of $\frac{\partial\widetilde{\mathbf{A}}^{\text{adi}}}{\partial P^{\alpha}}$)
includes an additional cross derivative force that indicates the difference
between these two equations. In other words, even though these two
QCLEs will propagate electrons equivalently, the difference in the
``effective'' force matrix will lead to different nuclear dynamics
in phase space. Specifically, from Eq.~\ref{eq:F-QCLE-WAD}, the
classical force on the $J$-th quasi-energy surface is given by $F_{JJ}^{\alpha}+\chi_{JJ}^{\alpha}$,
implying that the nuclear dynamics should experience the additional
cross derivative force on top of the Floquet quasi-energy surface.

\subsection{Adiabatic-then-Wigner (AW) approach}

Finally, for completeness, we should emphasize that the first derivation
of an F-QCLE was published by Horenko, Schmidt, and Schütte who used
an approach that was different from either of the approaches above.
In Ref.~\citep{horenko_theoretical_2001}, these authors derived
their F-QCLE using an Adiabatic-then-Wigner (AW) approach. Their final
equation of motion was:
\begin{eqnarray}
\frac{\partial}{\partial t}\widetilde{A}_{JK}^{\text{adi}} & = & -\frac{i}{\hbar}\left({\cal E}_{J}^{\text{F}}-{\cal E}_{K}^{\text{F}}\right)\widetilde{A}_{JK}^{\text{adi}}\nonumber \\
 &  & -\frac{P^{\alpha}}{M^{\alpha}}\frac{\partial\widetilde{A}_{JK}^{\text{adi}}}{\partial R^{\alpha}}-\frac{P^{\alpha}}{M^{\alpha}}\sum_{L}\left(D_{JL}^{\alpha}\widetilde{A}_{LK}^{\text{adi}}-\widetilde{A}_{JL}^{\text{adi}}D_{LK}^{\alpha}\right)\nonumber \\
 &  & +\frac{1}{2}\sum_{L}\left(\frac{\partial{\cal E}_{J}^{\text{F}}}{\partial R^{\alpha}}+\frac{\partial{\cal E}_{K}^{\text{F}}}{\partial R^{\alpha}}\right)\frac{\partial\widetilde{A}_{JK}^{\text{adi}}}{\partial P^{\alpha}}\label{eq:F-QCLE-AW}
\end{eqnarray}

As mentioned above, in the limit of a time-independent Hamiltonian,
Izmaylov and co-workers have shown that the AW approach results in
the loss of geometric phase related features for a 2D problem\citep{ryabinkin_analysis_2014};
the WA approach is far more accurate (see Figs in Ref.~\citep{ryabinkin_analysis_2014}).
Moreover, in contrast to Eq.~\eqref{eq:F-QCLE-WDA} and Eq.~\eqref{eq:F-QCLE-WAD},
the last term in Eq.~\ref{eq:F-QCLE-AW} does not include any off-diagonal
force term which implies no momentum rescaling. Within a surface hopping
context, a lack of momentum rescaling will inevitably lead to a violation
of detailed balance\citep{parandekar_detailed_2006,parandekar_mixed_2005,schmidt_mixed_2008}.
Note that early numerical assessments of the AW approach did not investigate
either geometric phase effects or detailed balance\citep{ando_non-adiabatic_2002,ando_mixed_2003}.
For these reasons, we expect the AW approach to the F-QCLE to be less
accurate than the WFA and WAF approaches. Nevertheless, to satisfy
the readers who is curious about the original AW approach, we will
present the relevant transmission and reflection probabilities and
final momentum distribution data in Appendix~\ref{sec:Horenko} which
should remind one why momentum jumps are so important for scattering
calculations\footnote{Interestingly, the AW approach to the time-independent QCLE was found
to perform well for short times for small closed systems in 2003\citep{ando_non-adiabatic_2002,ando_mixed_2003},
but we are unaware of more systematic tests of the method beyond Ref.~\citep{ryabinkin_analysis_2014},
especially within the context of long time dynamics.}.

\begin{table}
\begin{tabular}{c|c|c}
 & F-QCLE & effective force matrix\tabularnewline
\hline 
WFA & Eq.~\eqref{eq:F-QCLE-WDA} & $-\frac{\partial{\cal E}_{J}^{\text{F}}}{\partial R^{\alpha}}\delta_{JK}+({\cal E}_{J}^{\text{F}}-{\cal E}_{K}^{\text{F}})D_{JK}^{\alpha}$\tabularnewline
WAF & Eq.~\eqref{eq:F-QCLE-WAD} & $-\frac{\partial{\cal E}_{J}^{\text{F}}}{\partial R^{\alpha}}\delta_{JK}+({\cal E}_{J}^{\text{F}}-{\cal E}_{K}^{\text{F}})D_{JK}^{\alpha}-\sum_{\mu m}m\hbar\omega G_{\mu m}^{J*}\frac{\partial G_{\mu m}^{K}}{\partial R^{\alpha}}$\tabularnewline
AW & Eq.~\eqref{eq:F-QCLE-AW} & $-\frac{\partial{\cal E}_{J}^{\text{F}}}{\partial R^{\alpha}}\delta_{JK}$\tabularnewline
\end{tabular}

\caption{The F-QCLEs obtained via different approaches differ in the effective
force matrices for nuclear motions.\label{tab:F-QCLEs-via-different}}
\end{table}

\section{Results \label{sec:Results:}}

\subsection{Shifted avoided crossing model}

To analyze the difference between these approaches, we consider a
shifted avoided crossing model composed of a two-level electronic
system coupled to a 1D nuclear motion. The diabatic electronic states
are denoted as $|g\rangle$ and $|e\rangle$ and the electronic Hamiltonian
is given by 
\begin{equation}
\hat{V}(R,t)=\left(\begin{array}{cc}
V_{gg}(R) & V_{ge}(R,t)\\
V_{eg}(R,t) & V_{ee}(R)
\end{array}\right)
\end{equation}
where the diabatic energy is given by 
\begin{equation}
V_{gg}(R)=\begin{cases}
A(1-e^{-BR}) & R>0\\
-A(1-e^{BR}) & R<0
\end{cases}\label{eq:Vgg}
\end{equation}
\begin{equation}
V_{ee}(R)=-V_{gg}(R)+\hbar\omega
\end{equation}
and the diabatic coupling is periodic in time
\begin{equation}
V_{ge}(R,t)=V_{eg}(R,t)=Ce^{-DR^{2}}\cos\omega t.
\end{equation}
The parameters are $A=0.01$, \textbf{$B=1.6$}, $C=0.01$, $D=1.0$,
and the nuclear mass is $M=2000$. Note that this model is Tully\textquoteright s
simple avoided crossing model with two modifications: the diabatic
coupling becomes time-periodic and the excited potential energy surface
is shifted by $\hbar\omega$. For simplicity, we choose the laser
frequency $\hbar\omega=0.024$ large enough ($\hbar\omega>2A$) so
that the diabatic Floquet states $|gm\rangle$ have an avoided crossing
only with $|e(m-1)\rangle$ (at $R=0$) and do not have any trivial
crossings. The size of the Floquet basis is truncated by $m_{\text{max}}=4$
as the laser field is weak ($C/\hbar\omega<1$).

We assume the initial wavepacket is a Gaussian centered at the initial
position $R_{0}$ and momentum $P_{0}$ on diabat $|g\rangle$:
\begin{equation}
|\Psi_{0}\rangle=\frac{1}{{\cal N}}\exp\left(-\frac{(R-R_{0})^{2}}{2\sigma^{2}}+\frac{i}{\hbar}P_{0}(R-R_{0})\right)|g\rangle
\end{equation}
where the normalization factor is ${\cal N}^{4}=\pi\sigma^{2}$. The
width of the Gaussian is chosen to be $\sigma=20\hbar/P_{0}$. The
wavepacket can be propagated exactly in the diabatic representation.

\subsection{F-FSSH based on WFA and WAF}

To show the difference between the dynamics as obtained by the different
F-QCLEs, we will simulate F-FSSH results for both the WFA and WAF
approaches. Within F-FSSH, we describe the propagation of the Floquet
wavepacket by a swarm of trajectories, each with its own electronic
amplitudes $\tilde{c}_{J}$ satisfying 
\begin{equation}
\frac{\partial\tilde{c}_{J}}{\partial t}=-\frac{i}{\hbar}{\cal E}_{J}^{\text{F}}\tilde{c}_{J}-\frac{P}{M}\sum_{L}D_{JL}\tilde{c}_{L}
\end{equation}
where ${\cal E}_{J}^{\text{F}}={\cal E}_{J}^{\text{F}}(R)$, $D_{JL}=D_{JL}(R)$
and $R=R(t)$, $P=P(t)$ representing nuclear trajectory. All nuclear
trajectories move classically along an active Floquet state ($J$)
obeying
\begin{equation}
\frac{dR}{dt}=\frac{P}{M}
\end{equation}
\begin{equation}
\frac{dP}{dt}=\begin{cases}
-\frac{\partial{\cal E}_{J}^{\text{F}}}{\partial R} & \text{for WFA}\\
-\frac{\partial{\cal E}_{J}^{\text{F}}}{\partial R}-\chi_{JJ} & \text{for WAF}
\end{cases}\label{eq:dPdt}
\end{equation}
Here, based on the connection between the QCLE and FSSH, the nuclear
force in Eq.~\ref{eq:dPdt} is determined according to the diagonal
element of the effective force matrix in the F-QCLEs.

Consistent with the standard FSSH technique, the hopping probability
from active Floquet state $J$ to state $K$ is given by 
\[
\text{Prob}(J\rightarrow K)=-2\text{Re}\left(\frac{P^{\alpha}}{M^{\alpha}}\cdot D_{KJ}^{\alpha}\frac{\tilde{c}_{J}\tilde{c}_{K}^{*}}{|\tilde{c}_{J}|^{2}}\right)dt
\]
where $dt$ is the classical time step. After each successful hop,
the velocity is adjusted to conserve the total Floquet quasi-energy.
If a frustrated hop occurs, we implement velocity reversal\citep{jasper_improved_2003}.
Note that we neglect the decoherence correction since the over-coherence
problem should not be severe for a simple avoided crossing model\citep{landry_how_2012}.

In the end, the probability to measure diabatic state $\mu$ can be
evaluated by the density matrix interpretation\citep{landry_communication:_2013}
\begin{equation}
P_{\mu}=\sum_{m}\frac{N(\mu m)}{N_{\text{traj}}}+\sum_{n\neq m}\frac{\sum_{l}^{N(\mu m)}\sum_{k}^{N(\mu n)}\tilde{c}_{\mu m}^{(l)}\tilde{c}_{\mu n}^{(k)*}e^{i(m-n)\omega t}}{N(\mu m)N(\mu n)}
\end{equation}
where $N(\mu m)=\sum_{l}^{N_{\text{traj}}}\delta_{J^{(l)}\mu m}$
is the number of the trajectories that have the active surface $J^{(l)}$
end up on the Floquet state $|\mu m\rangle$. Here $l$ and $k$ are
the trajectory indices. We propagate $N_{\text{traj}}$ trajectories
for an amount of time long enough for each trajectory to pass through
the coupling region ($|R|<3$ for this parameter set).

\subsection{Effective Floquet quasi-energy surfaces for nuclear dynamics}

\begin{figure}
\includegraphics{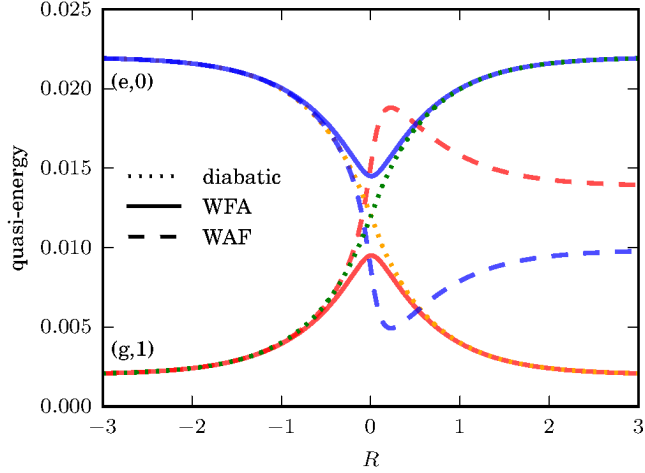}\caption{The effective potential energy surfaces for the WFA approach (solid
lines) and the WAF approach (dash lines). The lower (upper) quasi-energy
surface is in red (blue). The diabatic Floquet state energies $\widetilde{V}_{\mu m,\mu m}^{\text{F}}$
are plotted for $|g1\rangle$ (green) and $|e0\rangle$ (orange) in
dotted lines. Note that the barrier height and the equilibrium energy
ratio of the WAF surface is significantly modified (relative to the
WFA approach) by the presence of the additional cross derivative force.\label{fig:PES}}
\end{figure}

First, we analyze the effective potential energy surfaces for nuclear
dynamics by integrating Eq.~\ref{eq:dPdt} over $R$ for the WFA
and WAF approaches respectively. For the WFA approach, the effective
PES simply recovers the Floquet quasi-energy surfaces ${\cal E}_{J}^{\text{F}}$.
For the WAF approach, the effective PES is obtained by $V_{eff}(R)={\cal E}_{J}^{\text{F}}(R)+\int_{-\infty}^{R}\chi_{JJ}(R')dR'$
where the quasi-energy surface is modified by the integration of the
cross derivative force. We find that including the cross derivative
force in the WAF approach increases the crossing barrier for the nuclear
dynamics on the lower adiabat (see Fig.~\ref{fig:PES}). Note that,
in terms of an F-FSSH calculation, these changes will have a direct
effect on the nuclear dynamics, but not the electronic amplitudes.

\subsection{Transmission and reflection}

\begin{figure}
\includegraphics[bb=0bp 0bp 636bp 461bp,scale=0.45]{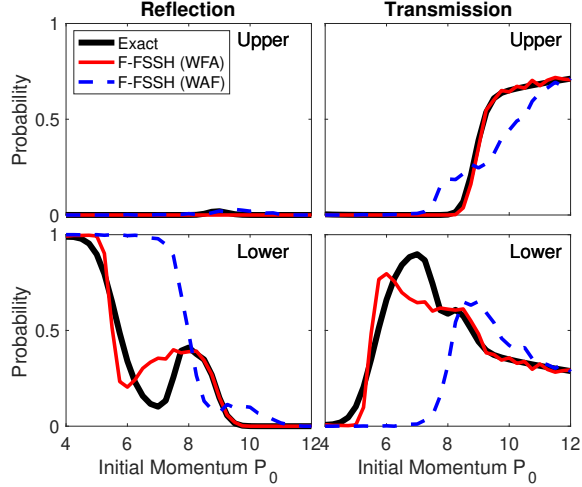}\caption{The probability of transmission (right) and reflection (left) on the
upper and lower adiabats as a function of the initial momentum. The
F-FSSH dynamics is implemented using the effective nuclear forces
of the WFA (red) and WAF (blue). Overall, the WFA result is more accurate
than the WAF result. The WFA result almost recover the correct nuclear
dynamics. Due to the cross derivative force, the nuclear trajectory
of the F-FSSH(WAF) experiences a much higher crossing barrier, requiring
larger $P_{0}$ for transmission. \label{fig:Transmission}}
\end{figure}

Next, we turn our attention to the transmission and reflection probabilities
produced by the F-FSSH calculations. Overall, the WFA results are
more accurate than the WAF results. In Fig.~\ref{fig:Transmission},
we find that, according to the WFA approach, there should be a rise
in transmission on the lower adiabat around $P_{0}\approx5.3$, which
is the momentum for which transmission should be allowed classically;
see the barrier height ($\approx0.007$) in Fig.~\ref{fig:PES}.
Indeed, such a threshold at $P_{0}\approx5.3$ is found according
to exact wavefunction simulation as well. However, for the WAF approach
with the cross derivative force, one find a higher crossing barrier
energy ($\approx0.015$), and the transmission on the lower adiabat
occurs (incorrectly) at $P_{0}\approx7.8$. This result suggests that
the cross derivative force is a fictitious term: the WAF semiclassical
derivation is spurious.

Let us now focus on the WFA results in more detail. Several points
are worth mentioning. First, the transmission to the upper adiabat
occurs after $P_{0}\approx8.9$, which agrees with the classical energy
difference $2A=0.02$ (see Eq.~\eqref{eq:Vgg} for the definition
of $A$). Second, for high initial momentum ($P_{0}>8.0$), the F-FSSH-WFA
can almost recover the correct nuclear dynamics. Third, in the intermediate
momentum region $P_{0}\in(6,8)$, unfortunately, the F-FSSH wavepacket
exhibits less transmission than the exact calculation. This discrepancy
may be attributed to FSSH's inability to capture nuclear tunneling
effects.

\section{Conclusion and Outlook\label{sec:Conclusion}}

We have analyzed three approaches for deriving the QCLE within a Floquet
formalism, and found three different F-QCLEs. While these F-QCLEs
take similar forms, the difference in the effective force matrix can
lead to large discrepancies in nuclear dynamics. As such, in the context
of driven electron-nuclear dynamics, our results reiterate the fact
that one cannot change the order of the operations in the derivation
of the correct QCLE. Specifically, as opposed to the WFA approach,
the WAF approach {[}exchanging Floquet electronic basis dressing (F)
and adiabatic transformation (A){]} is spurious. Overall, our results
are very consistent with the results of Izmaylov and Kapral\citep{ryabinkin_analysis_2014}
who find that one must be careful when deriving the QCLE even without
time dependence; they have already shown that the AW approach does
not include any geometric phase information. Thus, in the end, with
or without a time-dependent Hamiltonian, it appears that, provided
that one moves to the adiabatic representation \emph{as the very last
step}, one will always derive the correct semiclassical equation of
motion.

Looking forward, the derivation of the F-QCLE presented here validates
the F-FSSH method and paves the way to further improvements in the
future. With regard to coherence and decoherence, given the time-independent
nature of the Floquet Hamiltonian, we can immediately apply many decoherence
schemes, including augmented moment decoherence\citep{jain_efficient_2016,landry_how_2012,petit_how_2014,schwartz_quantum_1996,prezhdo_relationship_1998},
to the F-FSSH algorithm. As far as geometric phase effect is concerned,
it is known that Berry phases are already included within the QCLE\citep{subotnik_demonstration_2019}
for a time-independent electronic Hamiltonian, and so we would expect
that similar effects should already be included within this proper
F-QCLE for periodic (time-dependent) electronic Hamiltonians. Nevertheless,
however, there is one nuance which we have conveniently neglected
in the present paper. Note that, according to Eq.~\eqref{eq:double-bracket},
we have every reason to believe that the F-QCLE formalism (especially
for a non-monocrhomatic driving field with more than one Fourier mode
in the time-dependent Hamiltonian) will necessarily introduce a complex
(i.e. not real-valued) Floquet Hamiltonian. In such a case, we should
find not just Berry phases, but also Berry force\citep{miao_extension_2019}.
Future research into the nature of this intrinsic magnetic Berry force\textemdash how
or if it appears in the context of F-QCLE and surface hopping dynamics\textemdash is
currently underway and represents an exciting new direction for non-adiabatic
theory.

Lastly, it has been recently reported that novel control schemes,
such as Floquet engineering\citep{thanh_phuc_control_2018,schwennicke_optical_2020},
can enhance the excitation energy transfer rate even in the presence
of strong fluctuations and dissipation. Given so many potential applications
for F-QCLE simulations, we believe the present manuscript should find
immediate use in the physical chemistry and chemical physics community.

\section*{Acknowledgment}

This material is based upon work supported by the U.S. Department
of Energy, Office of Science, Office of Basic Energy Sciences under
Award Number DE-SC0019397. This research also used resources of the
National Energy Research Scientific Computing Center (NERSC), a U.S.
Department of Energy Office of Science User Facility operated under
Contract No. DE-AC02-05CH11231. We thank Abraham Nitzan for very helpful
discussions.

\section*{Data Availablity}

The data that support the findings of this study are available from
the corresponding author upon reasonable request.

\appendix

\section{Rewriting the Product of Wignerized Operators As Matrix Multiplication
in the Floquet Representation \label{sec:The-trick}}

Throughout this paper, we have constantly used one trick. Namely,
we have consistently rewritten the product of two Wigner transformed
operators (one of which must be time-periodic) into a non-standard
matrix product in the Floquet representation. To see how this trick
works in practice, we consider (for example) the operator product
$\hat{\rho}_{W}\hat{H}_{W}$.

We insert the electronic identity operator $\hat{1}=\sum_{\lambda}|\lambda\rangle\langle\lambda|$
in between $\hat{\rho}_{W}$ and $\hat{H}_{W}$: 
\begin{equation}
\langle\nu n|\hat{\rho}_{W}\hat{H}_{W}|\mu m\rangle=\sum_{\lambda}\langle\nu n|\hat{\rho}_{W}|\lambda\rangle\langle\lambda|\hat{H}_{W}|\mu m\rangle
\end{equation}
Next, as in Sec.~\ref{sec:Three-operations}, we express $\langle\lambda|\hat{H}_{W}|\mu m\rangle$
in the form of a Fourier series: 
\begin{equation}
\langle\lambda|\hat{H}_{W}|\mu m\rangle=\sum_{l}\langle\langle\lambda l|\hat{H}_{W}|\mu m\rangle\rangle e^{il\omega t}
\end{equation}
where the double bracket projection is defined by Fourier transform.
With the Fourier series, we write $|\lambda\rangle e^{il\omega t}=|\lambda l\rangle$
and obtain
\begin{equation}
\langle\nu n|\hat{\rho}_{W}\hat{H}_{W}|\mu m\rangle=\sum_{\lambda l}\langle\nu n|\hat{\rho}_{W}|\lambda l\rangle\langle\langle\lambda l|\hat{H}_{W}|\mu m\rangle\rangle
\end{equation}

\section{The AW approach\label{sec:Horenko}}

\begin{figure}
\includegraphics[scale=0.45]{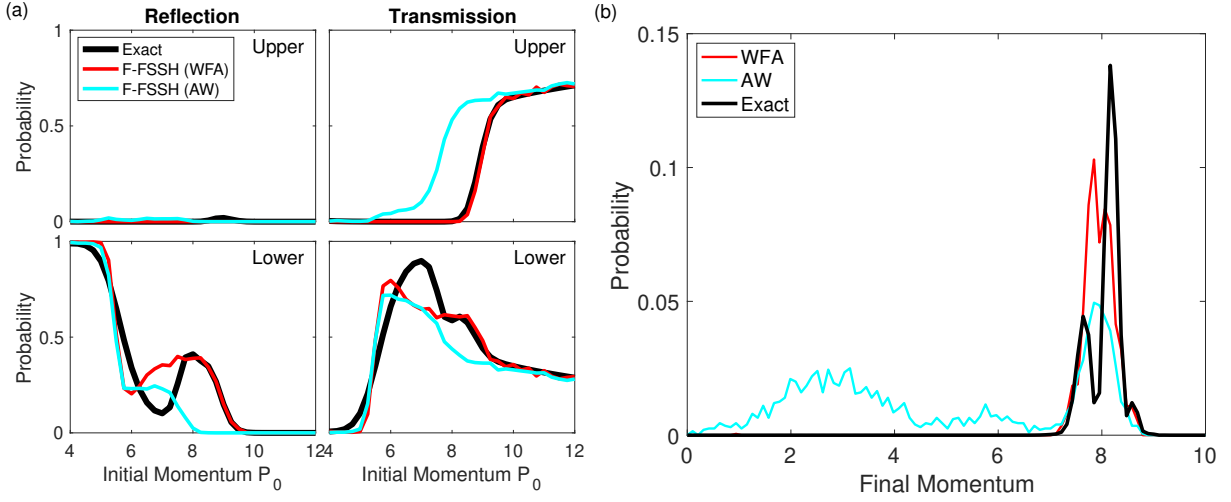}

\caption{(a) The transmission and reflection probabilities for the same scattering
problem as in Fig.~\ref{fig:Transmission} and (b) the final momentum
distribution (averaged over all channels) for incoming momentum $P_{0}=8$.
Here we calculate data using F-FSSH dynamics corresponding to the
WFA (red) and AW (cyan) approaches, aw well as exact wavepacket propagation
(black). The AW approach clearly leads to incorrect estimates of transmission
and reflection in the region $6\apprle P_{0}\apprle8$ due to the
lack of velocity rescaling, as well as an incorrect momentum distribution,
especially in the lower momentum region.\label{fig:momentum}}
\end{figure}
Following Ref.~\citep{horenko_theoretical_2001}, one can derive
the F-QCLE through an AW approach by invoking the wavefunction anstaz
$|\Psi(\vec{R},t)\rangle=\sum_{J}\tilde{a}_{J}(\vec{R},t)|\phi^{J}(\vec{R},t)\rangle$
in the Floquet adiabatic basis given by Eq.~\eqref{eq:Floquet_adiabatic_eigenstate},
then we insert this anstaz into the TDSE and obtain the equation of
motion for $\tilde{a}_{J}(\vec{R},t)$ in the form of $i\hbar\frac{\partial}{\partial t}\tilde{a}_{J}=\sum_{K}\hat{H}_{JK}^{\text{adi}}\tilde{a}_{K}$.
Here the adiabatic Hamiltonian operator is given by 
\begin{equation}
\hat{H}_{JK}^{\text{adi}}=\left({\cal E}_{J}^{\text{F}}(R^{\alpha})+\frac{(\hat{P}^{\alpha})^{2}}{2M^{\alpha}}\right)\delta_{JK}-i\hbar D_{JK}^{\alpha}\cdot\frac{\hat{P}^{\alpha}}{M^{\alpha}}+O(\hbar^{2})
\end{equation}
and the nuclear momentum operator is $\hat{P}^{\alpha}=-i\hbar\frac{\partial}{\partial R^{\alpha}}$.
Thereafter, one writes the equation of motion for the density matrix
($\widetilde{A}_{JK}^{\text{adi}}=\tilde{a}_{J}\tilde{a}_{K}^{*}$)
as $\frac{\partial}{\partial t}\widetilde{\mathbf{A}}^{\text{adi}}=-\frac{i}{\hbar}[\hat{\mathbf{H}}^{\text{adi}},\widetilde{\mathbf{A}}^{\text{adi}}]$
and performs a partial Wigner transformation using the Wigner\textendash Moyal
operator:
\begin{equation}
\frac{\partial}{\partial t}\widetilde{\mathbf{A}}^{\text{adi}}=-\frac{i}{\hbar}\left(\widetilde{\mathbf{H}}_{W}^{\text{adi}}e^{-i\hbar\overleftrightarrow{\Lambda}/2}\widetilde{\mathbf{A}}^{\text{adi}}-\widetilde{\mathbf{A}}^{\text{adi}}e^{-i\hbar\overleftrightarrow{\Lambda}/2}\widetilde{\mathbf{H}}_{W}^{\text{adi}}\right).
\end{equation}
The Wigner transform of the adiabatic Hamiltonian operator reads
\begin{equation}
(\widetilde{H}_{W}^{\text{adi}})_{JK}=\left({\cal E}_{J}^{\text{F}}(R^{\alpha})+\frac{(P^{\alpha})^{2}}{2M^{\alpha}}\right)\delta_{JK}-i\hbar D_{JK}^{\alpha}\cdot\frac{P^{\alpha}}{M^{\alpha}}.
\end{equation}
Next, following the same procedure in Sec.~II-A, we expand the Wigner\textendash Moyal
operator and truncate the equation of motion to first order of $\hbar$.
In the end, we obtain Eq.~\eqref{eq:F-QCLE-AW}, which is equivalent
to Eq.~(44) in Ref.~\citep{horenko_theoretical_2001} for the case
of the monochromatic external field (i.e. $\frac{dF_{0}}{dt}=0$ using
the notaion of Ref.~\citep{horenko_theoretical_2001}).

In Fig.~\ref{fig:momentum}, we compare the F-FSSH results for the
AW v.s. WFA F-QCLEs. Note that, because of the absence of the off-diagonal
force terms in Eq.~\eqref{eq:F-QCLE-AW}, the corresponding AW F-FSSH
trajectories neglect velocity rescaling when hopping between Floquet
adiabatic surfaces. Thus, an AW trajecory can hop to the upper adiabat
(the blue adiabatic potential in Fig.~\ref{fig:PES}) without worrying
about energy conservation. Note that, according to Fig.~\ref{fig:momentum},
the AW approach does outperform the WAF approach; see Fig.~\ref{fig:Transmission}.
However, the AW approach is far less accurate than the WFA approach.
In particular, Fig.~\ref{fig:momentum}(a) shows the AW approach
over-estimates the transmission probability on the upper adiabat (presumably
because of the lack of velocity rescaling). Furthermore, Fig.~\ref{fig:momentum}(b)
demostrates that the asymptotic velocities are also incorrect. Although
this data necessarily relies on an F-FSSH implementation, all together
Fig.~\ref{fig:momentum} strongly suggests that the AW approach to
the F-QCLE as given in Eq.~\eqref{eq:F-QCLE-AW} is not optimal\textemdash in
agreement with the conclusions of Izmaylov \emph{et al}.

\bibliographystyle{aipnum4-2}
\bibliography{ZoteroLibrary}

\end{document}